\documentclass[prb,onecolumn,showpacs,preprintnumbers,amsmath,amssymb,assume]{revtex4}
\usepackage{graphicx}		
\usepackage{dcolumn}				
\usepackage{bm}						
\usepackage{cases}
\usepackage{color}
\newcommand{\Bs}{&&\hspace{-3.2mm}} 

\begin{document}

\title{Christoffel equation in the polarization variables} 
\author{Vladimir Grechka$^{1}$}
\affiliation{$^{1}$Borehole Seismic, LLC}
\date{\today}

\begin{abstract}
	
We formulate the classic Christoffel equation in the polarization variables and solve it for the slowness vectors of plane waves corresponding to a given unit polarization vector. Our analysis shows that, unless the equation degenerates and yields an infinite number of different slowness vectors, the finite nonzero number of its legitimate solutions varies from 1 to 4. Also we find a subset of triclinic solids in which the polarization field can have holes; there exist finite-size solid angles of polarization directions unattainable to any plane wave.
	
\end{abstract}
\maketitle

\section{Introduction} 

The Christoffel \cite{Christoffel1877} equation governs the propagation of plane waves in homogeneous anisotropic media. A textbook approach \cite[e.g.,][]{Fedorov1968, Musgrave1970, Auld1973} to solving it consists in specifying the unit wavefront normal vector $\bm{n}$, constructing the symmetric, positive-definite Christoffel tensor $\bm{\Gamma}$, and computing solution to the ensuing eigenvalue-eigenvector problem~| the phase velocities $V$ and the unit polarization vectors $\bm{U}$ of three body waves propagating along the selected wavefront direction $\bm{n}$. If that $\bm{n}$ is not a singularity, all three vectors $\bm{U}$ are unique; alternatively, if $\bm{n}$ happens to be a singularity, some of vectors $\bm{U}$ or all of them are nonunique. 

The introduction of the slowness vector $\bm{p} = \bm{n}/V$ translates the uniqueness of $\bm{U}$ into uniqueness of the corresponding group-velocity vector $\bm{g}$ \cite[e.g.,][]{Fedorov1968, Musgrave1970}.
Its normalized version, the ray vector $\bm{r} \equiv \bm{g}/|\bm{g}|$, emerges in two-point ray-tracing \cite[e.g.,][]{Cerveny2001} as a natural variable for the description of velocities and polarizations of high-frequency waves, their number, for  a given ray direction $\bm{r}$ in a homogeneous anisotropic medium, ranging from three to nineteen \cite[][]{Grechka2017R2G}. Thus, all properties of plane waves propagating in a homogeneous anisotropic medium can be computed as functions of either unit vector $\bm{n}$ or $\bm{r}$. 

This paper discusses a similar computation for a unit polarization vector $\bm{U}$ selected as such a primary variable. A physical experiment encouraging the understanding of waves parameterized by $\bm{U}$ would be a record of wave motion by a single three-component sensor placed in a homogeneous elastic solid. If such a sensor is located sufficiently far from a source, it would record vectors $\bm{U}$ of individual body waves. Then one might wonder what kind of information about the local velocities or slownesses of these waves could be extracted from the measured direction of $\bm{U}$ and the knowledge of the medium~| the inquiry pursued here. 

\section{Theory}

An obvious point of departure for our investigation is the Christoffel equation \cite[e.g.,][]{Fedorov1968, Musgrave1970, Auld1973} 
\begin{equation} \label{eq:christ-03}
\bm{\Gamma}(\bm{p}) \bm{\cdot U} = \bm{U} \, ,
\end{equation}
where $\bm{p}$ is the slowness vector, $\bm{U}$ is the unit polarization vector,
\begin{equation} \label{eq:ani101-41}
\bm{\Gamma}(\bm{p}) \equiv  \bm{p \cdot c \cdot p} \,
\end{equation}
is the 
second-rank 
$3 \times 3$ positive-definite Christoffel tensor or matrix, and $\bm{c}$ is the 
fourth-rank 
$3 \times 3 \times 3 \times 3$ density-normalized stiffness tensor. We wish to solve equation~\ref{eq:christ-03} for $\bm{p}$ given the knowledge of $\bm{U}$ and $\bm{c}$.

One can immediately observe that function $\bm{p} = \bm{p}(\bm{U})$ is not necessarily unique. A simple example, exposing its nonuniqueness, is supplied by SH-waves propagating in the vertical $[\bm{x}_1, \, \bm{x}_3]$ plane of a purely isotropic medium, the waves that have identical polarization vectors $\bm{U}^\mathrm{SH} = [0, \, 1, \, 0]$ for any in-plane slowness vector $\bm{p}^\mathrm{\,SH} = \left[ p_1^\mathrm{\,SH}, \, 0, \, p_3^\mathrm{\,SH} \right]$. Still, the presented exception does not negate the fact that equation~\ref{eq:christ-03} is a system of three quadratic equations for the three components of slowness vector $\bm{p} = [\,p_1, \, p_2, \, p_3]$. When non-degenerative, system~\ref{eq:christ-03} has a finite number of roots, and B\'{e}zout's theorem \cite{Weisstein2003} equates their maximum number to the product of degrees of individual equations, that is, \mbox{$2 \times 2 \times 2 = 8$.} Also because a real-valued root $\bm{p}$ of system~\ref{eq:christ-03} is always accompanied by its centrally symmetric opposite $-\bm{p}$, \vspace{-2mm}
\begin{quote} 
	The maximum number of distinct (that is, non-centrally symmetric) real-valued roots of non-degenerative system~\ref{eq:christ-03} is equal to~4. 
\end{quote} \vspace{-2mm} 

Once these slowness roots are found, the corresponding phase and group velocities are given by \cite[][]{Fedorov1968, Musgrave1970, Auld1973}
\begin{equation} \label{eq:ani101-01}
V = \frac{1}{|\bm{p}|} \,
\end{equation}
and
\begin{equation} \label{eq:ani101-02}
\bm{g} = \bm{\Gamma}(\bm{U}) \bm{\cdot p} \, .
\end{equation}

\section{Isotropy} \label{ch:christ-iso}

The analysis of system~\ref{eq:christ-03}, shorthanded as the $\bm{p}(\bm{U})$ problem, is easiest in isotropic media, where the system does not even need to be solved to understand the properties of its solutions. Indeed, because $\bm{p}^\mathrm{\, P} \parallel \bm{U}^\mathrm{P}$ for the P-waves and $\bm{p}^\mathrm{\, S} \perp \bm{U}^\mathrm{S}$ for the S-waves, a given vector $\bm{U}$ uniquely constrains $\bm{p}^\mathrm{\, P}$ and yields infinitely many slowness vectors $\bm{p}^\mathrm{\, S}$, confined to the plane orthogonal to $\bm{U}$.

Let us derive this simple result from equation~\ref{eq:christ-03}. Because all directions are equivalent in isotropic media, the coordinate axis $\bm{x}_1$ can be oriented along our polarization vector $\bm{U}$ that becomes $\bm{U} = [1, \, 0, \, 0]$ in the new coordinate frame. Then, the substitution of that $\bm{U}$ in system~\ref{eq:christ-03}, written for isotropic stiffness tensor $\bm{c}$ expressed in terms of Lam\'{e} constants $\lambda$ and $\mu$, reduces the system to
\begin{subnumcases}{\label{eq:christ-14}}
(\lambda + 2 \, \mu) \, p_1^2 + \mu \, (p_2^2 + p_3^2) = 1 \, , & \\
\hspace{19.35mm} (\lambda + \mu) \, p_1 \, p_2 = 0 \, , & \\
\hspace{19.35mm} (\lambda + \mu) \, p_1 \, p_3 = 0 \, , &
\end{subnumcases}
implying two possible scenarios. 

\begin{description} \vspace{-2mm}
	\item[1.] If equations~\ref{eq:christ-14}b and~\ref{eq:christ-14}c  are satisfied by setting $p_2 = p_3 = 0$, equation~\ref{eq:christ-14}a yields a uniquely defined direction of the centrally symmetric P-wave slowness vector 
	\begin{equation} \label{eq:christ-15}
	\bm{p}^\mathrm{\, P} = \left[ \, \pm \frac{1}{\sqrt{\lambda + 2 \, \mu}} \, , 
	~ 0 \, , ~ 0 \, \right] \! .
	\end{equation}
	
	\item[2.] Alternatively, if both equations~\ref{eq:christ-14}b and~\ref{eq:christ-14}c are satisfied by setting $p_1 = 0$, equation~\ref{eq:christ-14}a, relating the two remaining unknowns $p_2$ and $p_3$, describes a circle in the $[\, p_2, \, p_3]$ plane, resulting in the S-wave slowness vectors
	\begin{equation} \label{eq:christ-16}
	\bm{p}^\mathrm{\, S} \equiv \bm{p}^\mathrm{\, S}(\varphi) = 
	\left[ \, 0 \, , ~ \frac{\sin \varphi}{\sqrt{\mu}} \, , 
	~ \frac{\cos \varphi}{\sqrt{\mu}} \, \right] ~ \forall ~ \varphi \, 
	\end{equation}
	and confirming the fact that the shear-wave slowness vector $\bm{p}^\mathrm{\, S}$ cannot be uniquely derived from polarization vector $\bm{U}$.
\end{description}  \vspace{-2mm}

\section{Vertical transverse isotropy} 

The $\bm{p}(\bm{U})$ problem in vertically transversely isotropic (VTI) media offers much more variety than that in just examined isotropic media, entailing several intriguing special cases. We will analyze them as they get discovered, starting from the general scenario, in which all three components of a given polarization vector $\bm{U}$ are nonzero, and system~\ref{eq:christ-03} reads \cite[][]{Fedorov1968, Tsvankin2001}
\begin{eqnarray} \label{eq:christ-17}
\Bs \left( \begin{array}{c c c}
c_{11} \, p_1^2 + c_{66} \, p_2^2 + c_{55} \, p_3^2 & (c_{11} - c_{66}) \, p_1 \, p_2 & 
(c_{13} + c_{55}) \, p_1 \, p_3 \\
(c_{11} - c_{66}) \, p_1 \, p_2 & c_{66} \, p_1^2 + c_{11} \, p_2^2 + c_{55} \, p_3^2 & 
(c_{13} + c_{55}) \, p_2 \, p_3 \\
(c_{13} + c_{55}) \, p_1 \, p_3 &  (c_{13} + c_{55}) \, p_2 \, p_3 & 
c_{55} \, (p_1^2 + p_2^2) + c_{33} \, p_3^2 \\
\end{array} \right) \!
 \bm{\cdot} \! \left[ \begin{array}{c} U_1 \\ U_2 \\ U_3 \end{array} \right] = 
\left[ \begin{array}{c} U_1 \\ U_2 \\ U_3 \end{array} \right] \! ,
\end{eqnarray}
where $c_{i j}$ $(i, \, j = 1, \, \ldots, \, 6)$ are the density normalized stiffness coefficients in Voigt notation.

\begin{figure}
	\centering
	\includegraphics[width=0.45\textwidth]{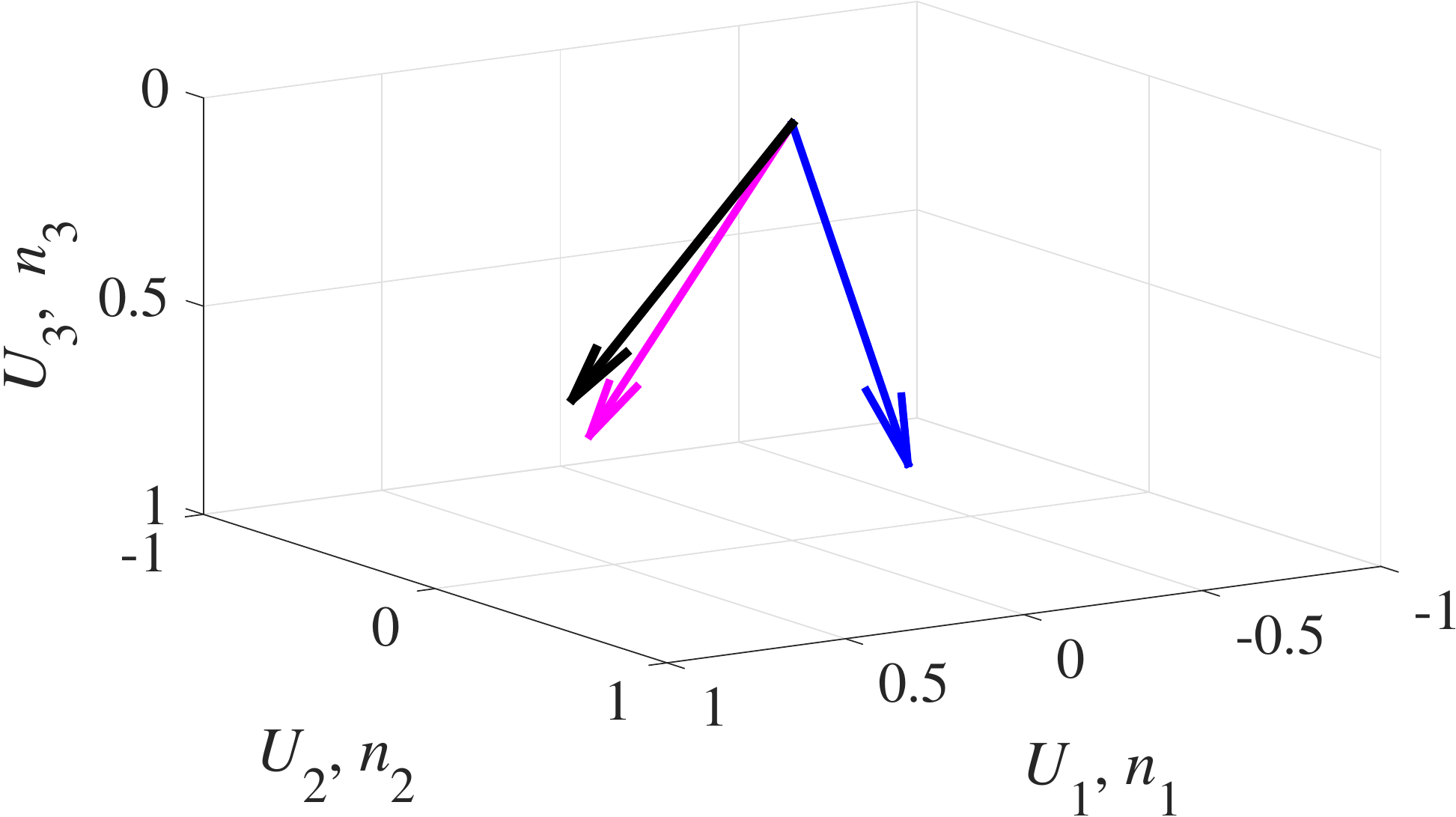}
	\caption{Input polarization vector $\bm{U} = [0.806, \, 0.293, \, 0.515]$ (the black arrow) and wavefront normals of the P- (the magenta arrow) and SV-waves (the blue arrow) calculated in VTI model~\ref{eq:christ-18} by solving equations~\ref{eq:christ-17}.
	}
	\label{fig:U2p-01}
\end{figure}

Mere inspection of matrix $\bm{\Gamma}(\bm{p})$ in the left side of system~\ref{eq:christ-17} reveals a special case: if $c_{13} = -c_{55}$ (unlikely for natural materials but mathematically possible), one eigenvector $\bm{U}$ of such a $\bm{\Gamma}(\bm{p})$ is always vertical, whereas two other eigenvectors are always horizontal regardless of the value of $\bm{p}$. Although superficial reaction to this observation might be that equations~\ref{eq:christ-17} become incompatible for an arbitrary direction of $\bm{U}$, a more careful investigation, presented below, is warranted.

When $c_{13} \neq -c_{55}$, as typically expected in VTI solids, matrix $\bm{\Gamma}(\bm{p})$ has one horizontal eigenvector $\bm{U} = \bm{U}^\mathrm{\,SH}$ corresponding to the SH-wave. Hence, unless $U_3 = 0$~| another special case~| the SH-wave slowness vector $\bm{p} = \bm{p}^\mathrm{\,SH}$ cannot be obtained by solving equations~\ref{eq:christ-17} for the components of $\bm{p}$. Consequently, equations~\ref{eq:christ-17} can have two real-valued roots that describe slowness vectors of the P- and SV-waves for an arbitrary polarization vector $\bm{U}$.

Figure~\ref{fig:U2p-01}, computed for a VTI stiffness matrix (in arbitrary units of velocity squared, as well as other stiffness matrixes below)
\begin{equation} \label{eq:christ-18}
\bm{c} = \left( \begin{array}{c c c c c c}
22.4 &  17.6 &  9.5 & 0 & 0 & 0 \\
~ & 22.4 &  9.5 & 0 & 0 & 0 \\
~ &  ~ & 16 & 0 & 0 & 0 \\
~ & \mathrm{SYM} & ~ & 4 & 0 & 0 \\
~ &  ~ &  ~ &           ~ & 4 & 0 \\
~ &  ~ &  ~ &           ~ & ~ & 2.4 \\
\end{array} \right) \! ,
\end{equation}
with ``SYM'' denoting the symmetric part of $\bm{c}$, illustrates this arrangement. Equations~\ref{eq:christ-17} have the real-valued slowness solutions $\bm{p} = \bm{p}^\mathrm{\,P}$ and $\bm{p} = \bm{p}^\mathrm{\,SV}$, and Figure~\ref{fig:U2p-01} displays the corresponding wavefront normals $\bm{n} = \bm{p}/|\bm{p}|$. The P-wave normal $\bm{n}^\mathrm{P}$ (magenta) is close to the polarization vector $\bm{U}$ (black), whereas the SV normal $\bm{n}^\mathrm{SV}$ (blue) is approximately orthogonal to it. 

The two slowness solutions shown in Figure~\ref{fig:U2p-01}, however, do not exhaust all the possibilities. If a VTI model possesses intersection singularities \cite[in the terminology of][]{CrampinYedlin1981} $\bm{n}^s$, polarization vector $\bm{U}(\bm{n}^s)$ at a singularity might happen to be equal to an input vector $\bm{U}$, making the singular slowness vector $\bm{p}^s$ a part of real-valued solution of equations~\ref{eq:christ-17}; also because a vertical plane containing an arbitrary vector $\bm{U}$ is a symmetry plane of a VTI medium, singular solutions $\bm{p}^s$ of equations~\ref{eq:christ-17} always come in non-centrally symmetric pairs.

\begin{figure}
	\centering
	\includegraphics[width=0.45\textwidth]{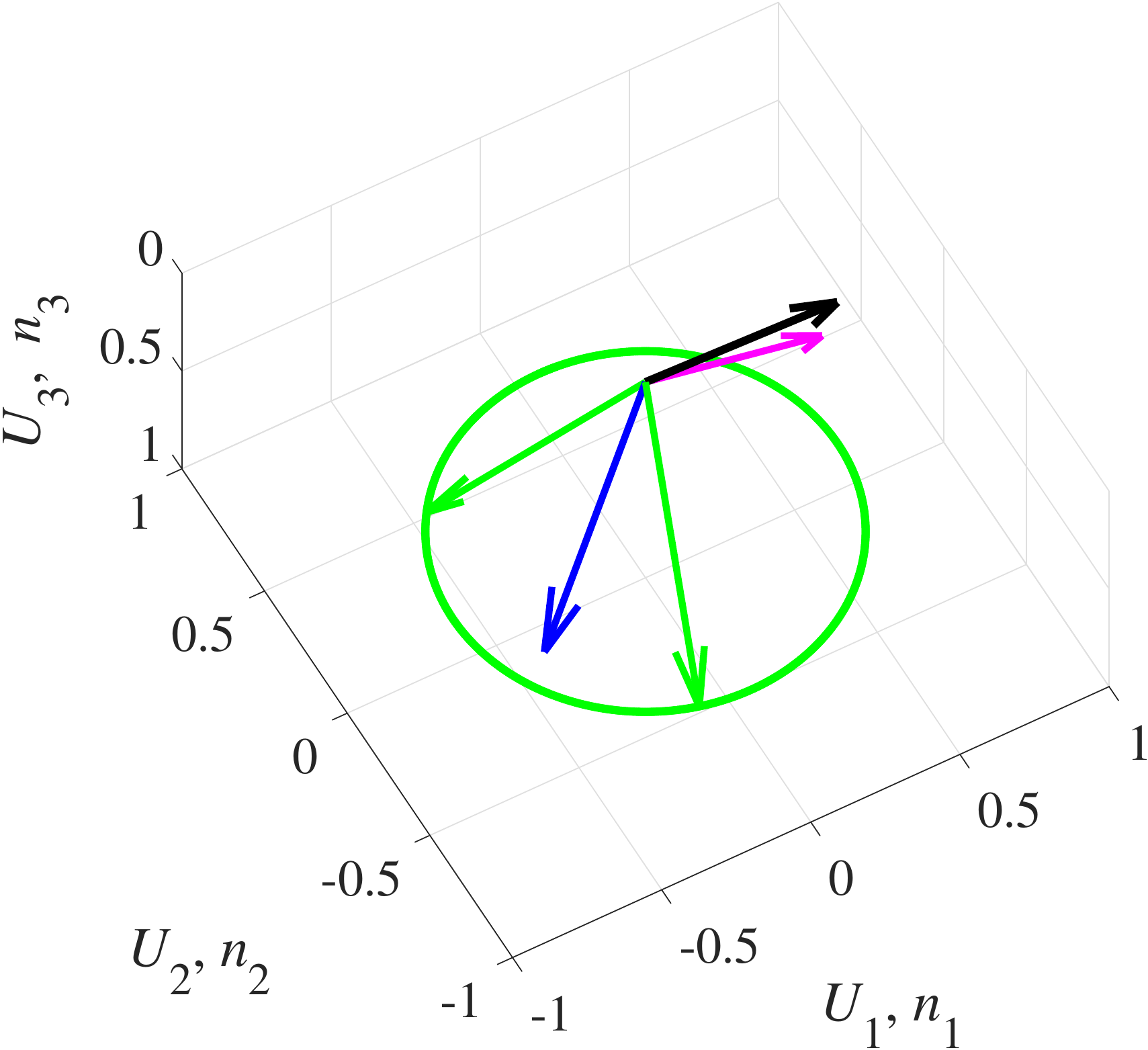}
	\caption{Same as Figure~\ref{fig:U2p-01} but for VTI model~\ref{eq:christ-19}. In addition to wavefront normals of the P- (the magenta arrow) and SV-waves (the blue arrow), solutions of equations~\ref{eq:christ-17} contain two singular normals (the green arrows), connected to the green circle that marks the SV-SH wave intersection singularity.
	}
	\label{fig:U2p-02}
\end{figure}

To investigate this scenario, we increase the stiffness coefficient $c_{66}$ in matrix~\ref{eq:christ-18} to make $c_{66} > c_{55}$ and create an intersection singularity in VTI model
\begin{equation} \label{eq:christ-19}
\bm{c} = \left( \begin{array}{c c c c c c}
22.4 &  11.2 &  9.5 & 0 & 0 & 0 \\
~ & 22.4 &  9.5 & 0 & 0 & 0 \\
~ &  ~ & 16 & 0 & 0 & 0 \\
~ & \mathrm{SYM} & ~ & 4 & 0 & 0 \\
~ &  ~ &  ~ &           ~ & 4 & 0 \\
~ &  ~ &  ~ &           ~ & ~ & 5.6 \\
\end{array} \right) \! 
\end{equation}
at $\bm{n}^s = [0.656, \, 0, \, 0.764]$ in the $[\bm{x}_1, \, \bm{x}_3]$ plane. Figure~\ref{fig:U2p-02} presents the obtained wavefront normals for the same polarization vector $\bm{U} = [0.806, \, 0.293, \, 0.515]$ (the black arrow) as that in Figure~\ref{fig:U2p-01}. Indeed, the two singular wavefront normals $\bm{n}^s$ (the green arrows in Figure~\ref{fig:U2p-02}) have been recovered from equations~\ref{eq:christ-17} in addition to $\bm{n}^\mathrm{P}$ (the magenta arrow) and $\bm{n}^\mathrm{SV}$ (the blue arrow).

\subsection{Equality $\bm{c_{13} = -c_{55}}$}

The presented example allows us to understand how to properly treat the special case of equality
\begin{equation} \label{eq:christ-20}
c_{13} = -c_{55} \, .
\end{equation}
Even though matrix $\bm{\Gamma}(\bm{p})$ in system~\ref{eq:christ-17} does have one vertical and two horizontal eigenvectors, the presence of intersection singularities could make equations~\ref{eq:christ-17} compatible for an arbitrary polarization vector $\bm{U}$. To illustrate that, let us impose constraint~\ref{eq:christ-20} on VTI stiffness matrix~\ref{eq:christ-19}, so that it becomes
\begin{equation} \label{eq:christ-21}
\bm{c} = \left( \begin{array}{c c r c c c}
22.4 &  11.2 &  -4 & 0 & 0 & 0 \\
~ & 22.4 &  -4 & 0 & 0 & 0 \\
~ &  ~ & 16 & 0 & 0 & 0 \\
~ & \mathrm{SYM} & ~ & 4 & 0 & 0 \\
~ &  ~ &  ~ &           ~ & 4 & 0 \\
~ &  ~ &  ~ &           ~ & ~ & 5.6 \\
\end{array} \right) \! .
\end{equation}

\begin{figure}
	\centering
	\begin{tabular}{c c}
		~ \qquad (a) &	~ \qquad (b) \\
		\includegraphics[width=0.4\textwidth]{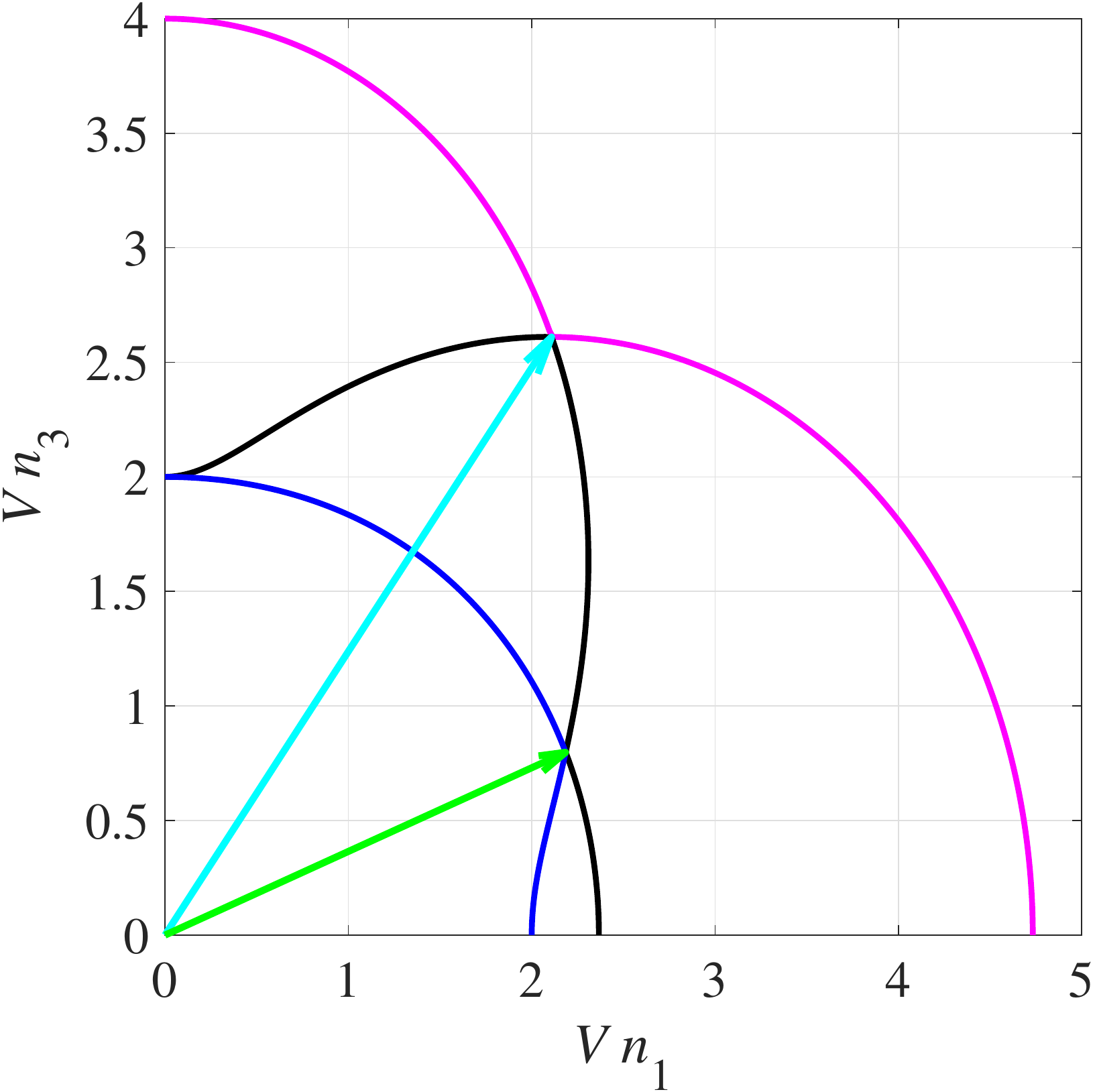} &
		\includegraphics[width=0.4\textwidth]{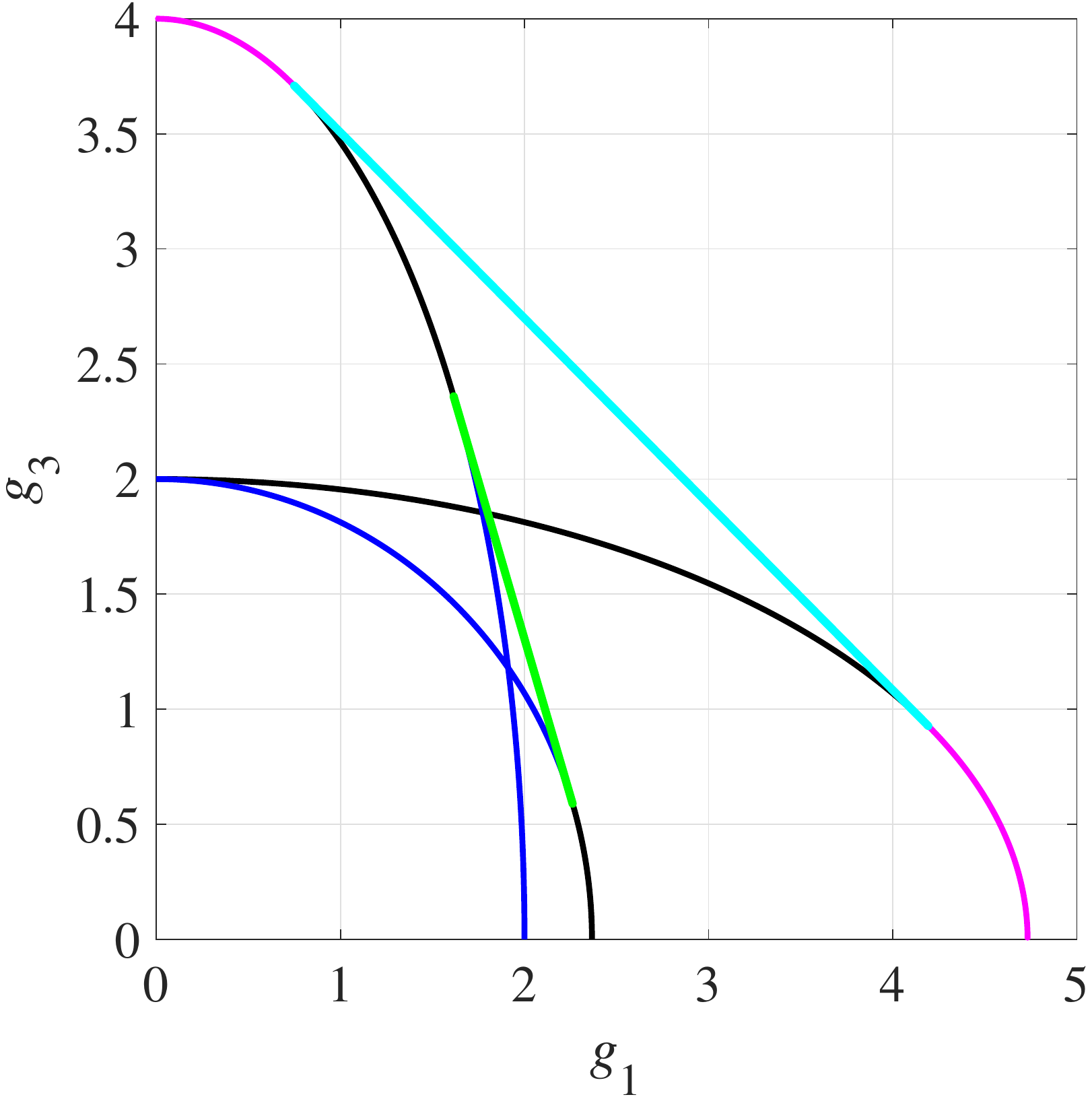} 
	\end{tabular}
	\caption{Quadrants of (a) phase- and (b) group-velocity surfaces of the P- (magenta), S$_1$- (black) and S$_2$-waves (blue) in the vertical plane $[\bm{x}_1, \, \bm{x}_3]$ of VTI model~\ref{eq:christ-21}. The directions of two intersection singularities $\bm{n}^{s1} = [0.628, \, 0, \, 0.778]$ and $\bm{n}^{s2} = [0.939, \, 0, \, 0.343]$ are marked by the cyan and green arrows, respectively, in (a), the corresponding degenerative internal refraction cones~| by the cyan and green lines in (b).
	}
	\label{fig:U2p-03}
\end{figure}

\begin{figure}
	\centering
	\includegraphics[width=0.45\textwidth]{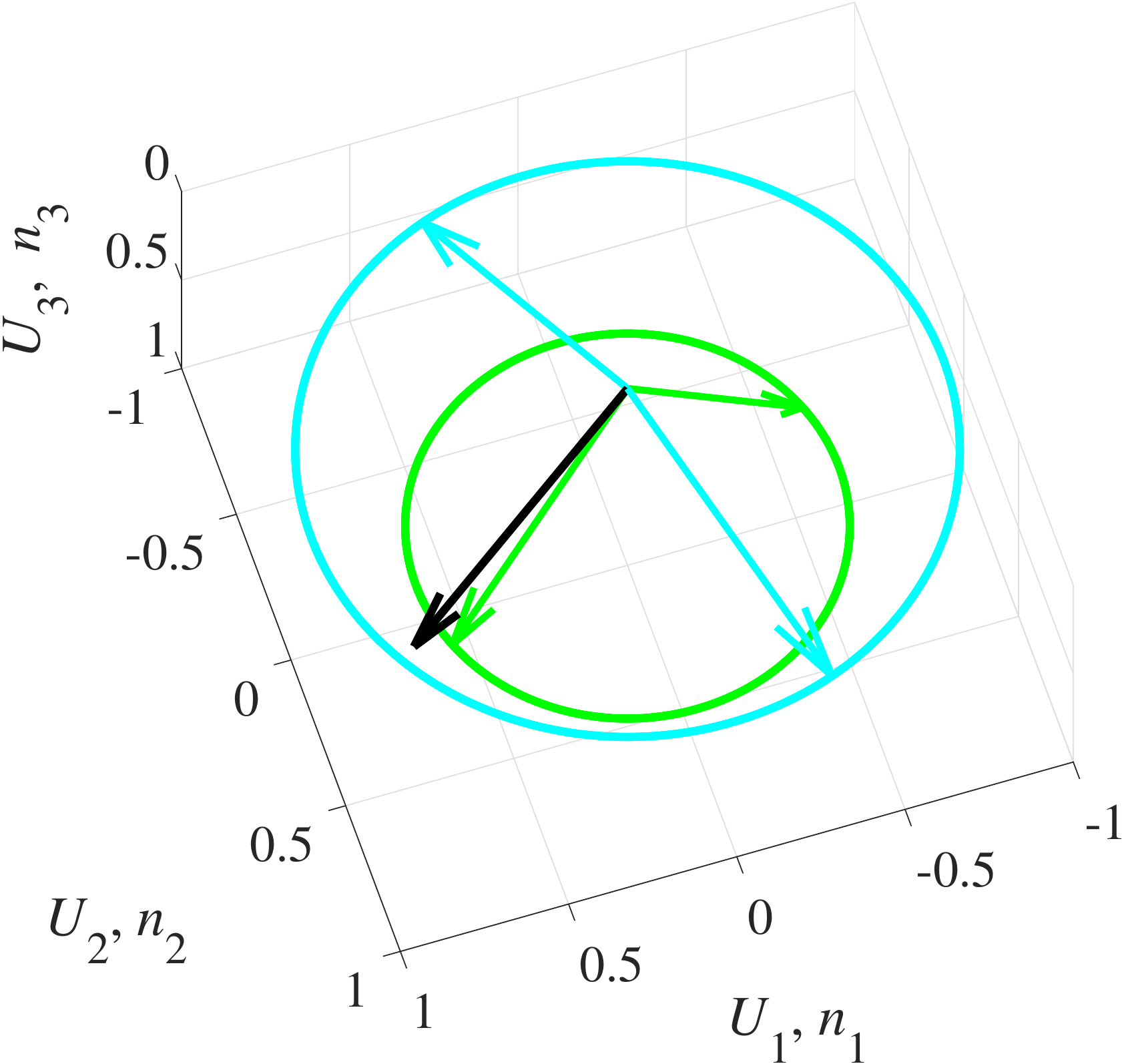}
	\caption{Same as Figure~\ref{fig:U2p-02} but for VTI model~\ref{eq:christ-21}. All four solutions of equations~\ref{eq:christ-17}, marked by the green and cyan arrows, for polarization vector $\bm{U}$ (the black arrow) correspond to intersection singularities indicated by the circles.
	}
	\label{fig:U2p-04}
\end{figure}

Model~\ref{eq:christ-21} exhibits two intersection singularities $\bm{n}^{s1}$ and $\bm{n}^{s2}$, indicated by the cyan and green arrows in Figure~\ref{fig:U2p-03}a, and two corresponding internal refraction cones, degenerating into the cyan and green straight lines in Figure~\ref{fig:U2p-03}b. The fans of polarization vectors $\bm{U}(\bm{n}^{s1})$ and $\bm{U}(\bm{n}^{s2})$ at those singularities contain vectors equal to the input polarization vector $\bm{U} = [0.720, \, 0.262, \,  0.643]$ (the black arrow in Figure~\ref{fig:U2p-04}). Hence, all four solutions of equations~\ref{eq:christ-17} for this vector $\bm{U}$ come from the intersection singularities.

Finally, let us investigate two symmetric orientations of vector $\bm{U}$~| along the vertical symmetry axis $\mathbf{a}$ and orthogonally to it.


\subsection{Vertical polarization vector}

When input polarization vector is vertical, $\bm{U} = [0, \, 0, \, 1]$, system~\ref{eq:christ-17} simplifies to  
\begin{subnumcases}{\label{eq:christ-22}}
\hspace{7.5mm} (c_{13} + c_{55}) \, p_1 \, p_3 = 0 \, , & \\
\hspace{7.5mm} (c_{13} + c_{55}) \, p_2 \, p_3 = 0 \, , & \\		
c_{55} \, (p_1^2 + p_2^2) + c_{33} \, p_3^2 = 1 & 
\end{subnumcases}
and becomes very similar to system~\ref{eq:christ-14} examined for isotropic media. Consequently, three types of solutions are possible~| two already analyzed in section~\ref{ch:christ-iso} and one related to the equality $c_{13} + c_{55} = 0$, whose analog $\lambda + \mu = 0$ is prohibited for isotropy by the elastic stability conditions.

\begin{description} \vspace{-2mm}
	\item[1.] When equations~\ref{eq:christ-22}a and~\ref{eq:christ-22}b are satisfied by setting $p_1 = p_2 = 0$, equation~\ref{eq:christ-22}c yields the vertical P-wave slowness vector
	\begin{equation} \label{eq:christ-23}
	\bm{p}^\mathrm{\, P} = \left[ \, 0 \, , ~ 0 \, , ~ \pm \frac{1}{\sqrt{c_{33}}} \, \right] \! .
	\end{equation}
	
	\item[2.] When equations~\ref{eq:christ-22}a and~\ref{eq:christ-22}b are satisfied by setting $p_3 = 0$, equation~\ref{eq:christ-22}c, relating the two remaining unknowns $p_1$ and $p_2$, describes infinitely many shear-wave slowness vectors
	\begin{equation} \label{eq:christ-24}
	\bm{p}^\mathrm{\, S} \equiv \bm{p}^\mathrm{\, S}(\varphi) = 
	\left[ \, \frac{\sin \varphi}{\sqrt{c_{55}}} \, , 
	~ \frac{\cos \varphi}{\sqrt{c_{55}}} \, , ~ 0 \, \right] ~ \forall ~ \varphi \, ,
	\end{equation}
	their ends tracing a circle with radius $1/\sqrt{c_{55}}$ in the $[\, p_1, \, p_2]$ plane.
	
	\item[3.] Finally, when $c_{13} + c_{55} = 0$, equations~\ref{eq:christ-22}a and~\ref{eq:christ-22}b are satisfied identically for any slowness vector, whereas  equation~\ref{eq:christ-22}c constraints the components of $\bm{p}$ to the surface of a spheroid in the $\bm{p}$-space. 
\end{description} \vspace{-2mm}

\subsection{Horizontal polarization vector}

Because all horizontal directions in a VTI medium are equivalent due to its rotational invariance around the vertical, let us select $\bm{U} = [1, \, 0, \, 0]$ to simplify our analysis. System~\ref{eq:christ-17} then reads
\begin{subnumcases}{\label{eq:christ-25}}
c_{11} \, p_1^2 + c_{66} \, p_2^2 + c_{55} \, p_3^2 = 1 \, , & \\
\hspace{10mm} (c_{11} - c_{66}) \, p_1 \, p_2 = 0 \, , & \\		
\hspace{10mm} (c_{13} + c_{55}) \, p_1 \, p_3 = 0 \, , & 
\end{subnumcases}
implying three possibilities similar to those just discussed for the vertical vector~$\bm{U}$. 

\begin{description} \vspace{-2mm}
	\item[1.] When equations~\ref{eq:christ-25}b and~\ref{eq:christ-25}c are satisfied for $p_2 = p_3 = 0$, equation~\ref{eq:christ-25}a describes the horizontal P-wave slowness vector
	\begin{equation} \label{eq:christ-26}
	\bm{p}^\mathrm{\, P} = \left[ \, \pm\frac{1}{\sqrt{c_{11}}} \, , ~ 0 \, , ~ 0 \, \right] \! .
	\end{equation}
	
	\item[2.] When both equations~\ref{eq:christ-25}b and~\ref{eq:christ-25}c are satisfied for $p_1 = 0$, equation~\ref{eq:christ-25}a yields infinitely many SH-wave slowness vectors
	\begin{equation} \label{eq:christ-27}
	\bm{p}^\mathrm{\, SH} \equiv \bm{p}^\mathrm{\, SH}(\varphi) = 
	\left[ \, 0 \, , ~\frac{\sin \varphi}{\sqrt{c_{66}}} \, , 
	~ \frac{\cos \varphi}{\sqrt{c_{55}}} \, \right] ~ \forall ~ \varphi \, ,
	\end{equation}
	their ends placed at an ellipse in the vertical $[\, p_2, \, p_3]$ plane.
	
	\item[3.] When $c_{13} + c_{55} = 0$ and equation~\ref{eq:christ-25}b is satisfied for $p_1 = 0$ ($c_{11} - c_{66} > 0$ in accordance with the elastic stability conditions in VTI media),  equation~\ref{eq:christ-25}a describes the slowness vectors given by equation~\ref{eq:christ-27}. Alternatively, when $c_{13} + c_{55} = 0$ but equation~\ref{eq:christ-25}b is satisfied for $p_2 = 0$ instead of $p_1 = 0$, equation~\ref{eq:christ-25}a describes a set of the slowness vectors in the $[\, p_1, \, p_3]$ plane,
	\begin{equation} \label{eq:christ-28}
	\bm{p} = \left[ \, \frac{\sin \varphi}{\sqrt{c_{11}}} \, , ~ 0 \, , ~ 
	\frac{\cos \varphi}{\sqrt{c_{55}}} \, \right] ~ \forall ~ \varphi \, .
	\end{equation}
\end{description} \vspace{-2mm}

\section{Symmetries lower than transverse isotropy} \label{ch:christ-tri}

\begin{figure}
	\centering
	\includegraphics[width=0.45\textwidth]{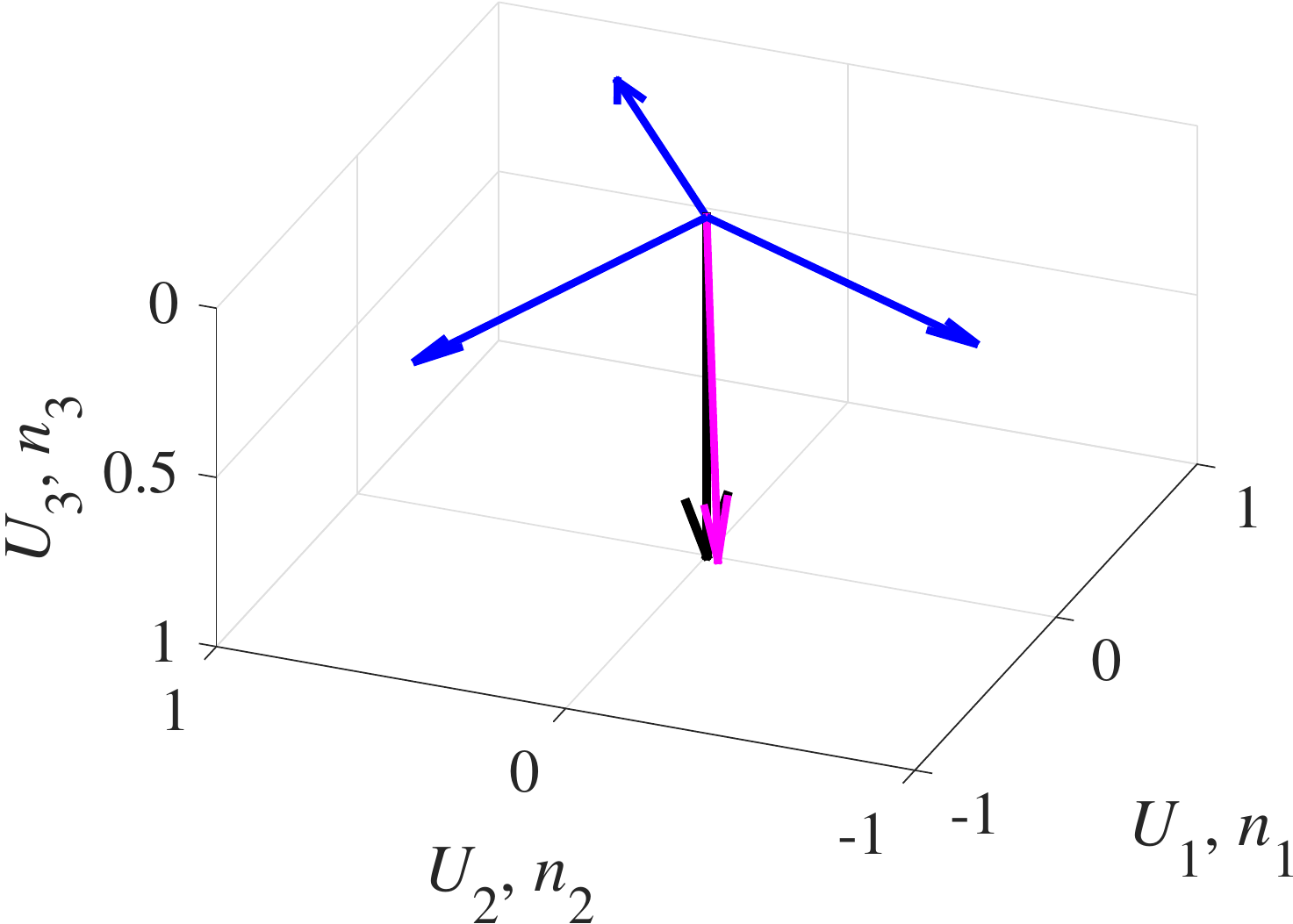}
	\caption{Same as Figure~\ref{fig:U2p-01} but for triclinic model~\ref{eq:christ-29}. The input vertical polarization vector $\bm{U} = [0, \, 0, \, 1]$ is indicated by the black arrow, the wavefront normals of one P-wave and three S-waves~| by the magenta and blue arrows, respectively.
	}
	\label{fig:U2p-05}
\end{figure}

The analysis of the $\bm{p}(\bm{U})$ problem presented so far reveals that system~\ref{eq:christ-03} can have two, four, or infinitely many slowness solutions for a given polarization vector $\bm{U}$. Clearly, the first two possibilities are realizable in generally anisotropic media because roots of a polynomial system are continuous functions of its coefficients, which are, in turn, continuous functions of the stiffness components. For instance, Figure~\ref{fig:U2p-05} displays four singularity-unrelated solutions (the magenta and blue arrows) for triclinic model
\begin{equation} \label{eq:christ-29}
\bm{c} = \left( \begin{array}{c c c r r r}
50 & 10 & 20 & -5 & 0 & 10 \\
~ &  50 & 30 &  3 & -5 & 10 \\
~ &  ~  & 50 &  2 & 1 & 0 \\
~ & \mathrm{SYM} & ~ & 20 & 3 & 5 \\
~ &  ~ &  ~ &           ~ & 30 & -5 \\
~ &  ~ &  ~ &           ~ & ~ & 10 \\
\end{array} \right) \! .
\end{equation}

The number of real-valued roots of equations~\ref{eq:christ-03} in low-symmetry anisotropic media can be infinite, too. Consider, for example, the vertical polarization vector $\bm{U} = [0, \, 0, \, 1]$ in an orthorhombic solid described by its generic stiffness matrix. 
There, equations~\ref{eq:christ-03} reduce to a system similar to~\ref{eq:christ-22},
\begin{subnumcases}{\label{eq:christ-28.1}}
\hspace{9.65mm} (c_{13} + c_{55}) \, p_1 \, p_3 = 0 \, , & \\
\hspace{9.65mm} (c_{23} + c_{44}) \, p_2 \, p_3 = 0 \, , & \\		
c_{55} \, p_1^2 + c_{44} \, p_2^2 + c_{33} \, p_3^2 = 1 \, , & 
\end{subnumcases}
and because equations~\ref{eq:christ-28.1}a and~\ref{eq:christ-28.1}b are simultaneously satisfied at $p_3 = 0$, equation~\ref{eq:christ-28.1}c yields an infinite number of slowness vectors
\begin{equation} \label{eq:christ-28.2}
\bm{p} = \left[ \, \frac{\sin \varphi}{\sqrt{c_{55}}} \, , ~ 
\frac{\cos \varphi}{\sqrt{c_{44}}} \, , ~ 0 \, \right] ~ \forall ~ \varphi \, .
\end{equation}

Hence, it remains to investigate whether equations~\ref{eq:christ-03} can have zero, one, or three real-valued roots. To address the question pertaining to the existence of one or three real-valued roots in a systematic way, we select a local coordinate frame, in which our input polarization vector $\bm{U} = [0, \, 0, \, 1]$, and explicitly write system~\ref{eq:christ-03} for that $\bm{U}$ in triclinic media. The system reads
\begin{subnumcases}{\label{eq:christ-33}}
c_{1 5} \, p_1^2 + (c_{1 4} + c_{5 6}) \, p_1 \, p_2 + (c_{1 3} + c_{5 5}) \, p_1 \, p_3 + 
c_{4 6} \, p_2^2 \, & \nonumber \\
\hspace{9.2mm} + \, (c_{3 6} + c_{4 5}) \, p_2 \, p_3 + c_{3 5} \, p_3^2 = 0 \, , & \\
c_{5 6} \, p_1^2 + (c_{2 5} + c_{4 6}) \, p_1 \, p_2 + (c_{3 6} + c_{4 5}) \, p_1 \, p_3 + 
c_{2 4} \, p_2^2 \, & \nonumber \\
\hspace{9.2mm} + \, (c_{2 3} + c_{4 4}) \, p_2 \, p_3 + c_{3 4} \, p_3^2 = 0 \, , & \\
c_{5 5} \, p_1^2 + 2 \, c_{4 5} \, p_1 \, p_2 + 2 \, c_{3 5} \, p_1 \, p_3 + 
c_{4 4} \, p_2^2 + 2 \, c_{3 4} \, p_2 \, p_3 + c_{3 3} \, p_3^2 = 1 \, , &  
\end{subnumcases}
lending itself to straightforward analysis.   

\subsection{One root}

If a triclinic model is such that 
\begin{subnumcases}{\label{eq:christ-34}}
c_{1 4} + c_{5 6} = 0 \, , & \\
c_{1 3} + c_{5 5} = 0 \, , & \\
c_{3 6} + c_{4 5} = 0 \, , &
\end{subnumcases}
the cross-terms vanish in equation~\ref{eq:christ-33}a; if additionally  
\begin{equation} \label{eq:christ-35}
c_{3 4} = c_{3 5} = 0 \, ,
\end{equation}
the terms proportional to $p_3^2$ vanish in both equations~\ref{eq:christ-33}a or~\ref{eq:christ-33}b; finally, if the product
\begin{equation} \label{eq:christ-36}
c_{1 5} \, c_{4 6} > 0 \, ,
\end{equation}
equations~\ref{eq:christ-33}a and~\ref{eq:christ-33}b are satisfied for a single pair of real-valued slowness components $p_1 = p_2 = 0$, and equation~\ref{eq:christ-33}c yields a single centrally symmetric slowness root
\begin{equation} \label{eq:christ-37}
\bm{p}^\mathrm{P} = \left[ \, 0 \, , ~ 0 \, , ~ \pm \frac{1}{\sqrt{c_{33}}} \, \right] .
\end{equation}

\subsection{Three roots}

\begin{figure}
	\centering
	\includegraphics[width=0.45\textwidth]{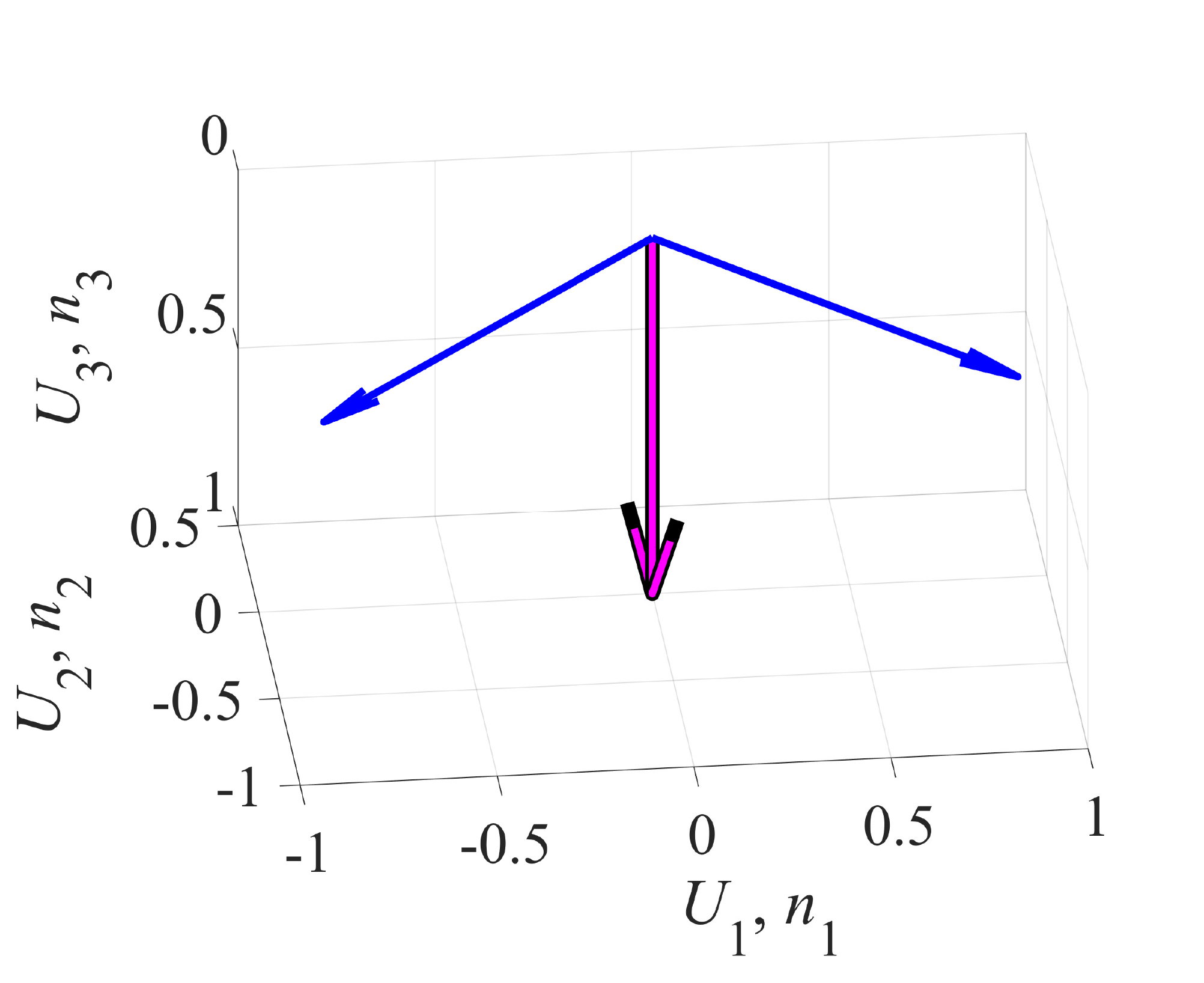}
	\caption{Same as Figure~\ref{fig:U2p-05} but for model~\ref{eq:christ-39.1}. 
	}
	\label{fig:U2p-06.1}
\end{figure}

Alternatively, if the opposite of inequality~\ref{eq:christ-36} is true,
\begin{equation} \label{eq:christ-38}
c_{1 5} \, c_{4 6} < 0 \, ,
\end{equation}
equation~\ref{eq:christ-33}a has two roots, related to each other as
\begin{equation} \label{eq:christ-39}
\frac{p_1}{p_2}  = \pm \, \sqrt{- \frac{c_{4 6}}{c_{1 5}}}  \,  
\end{equation}
and complementing the already discussed root $p_1 = p_2 = 0$. Therefore, system~\ref{eq:christ-33} can possess three real-valued roots, as shown in Figure~\ref{fig:U2p-06.1} for triclinic stiffness matrix
\begin{equation} \label{eq:christ-39.1}
\bm{c} = \left( \begin{array}{c c r r r r}
50 & 10 & -20 & -5 & 1 & 10 \\
~ &  50 &  20 &  3 & 6 & 10 \\
~ &  ~  &  50 &  0 & 0 & 2 \\
~ & \mathrm{SYM} & ~ & 22 & -2 & -5 \\
~ &  ~ &  ~ &           ~ & 20 & 5 \\
~ &  ~ &  ~ &           ~ & ~ & 10 \\
\end{array} \right) \! .
\end{equation}

The logic presented for equation~\ref{eq:christ-33}a would apply to equation~\ref{eq:christ-33}b when the set of equalities~\ref{eq:christ-34} is replaced by
\begin{subnumcases}{\label{eq:christ-40}}
c_{2 5} + c_{4 6} = 0 \, , & \\
c_{3 6} + c_{4 5} = 0 \, , & \\
c_{2 3} + c_{4 4} = 0 \, , &
\end{subnumcases}
and inequalities~\ref{eq:christ-36} or~\ref{eq:christ-38} are replaced by
\begin{equation} \label{eq:christ-41}
c_{5 6} \, c_{2 4} > 0 \, 
\end{equation}
or
\begin{equation} \label{eq:christ-42}
c_{5 6} \, c_{2 4} < 0 \, ,
\end{equation}
respectively.

\section{Prohibited polarization directions} 

An unexpected scenario arises when equations~\ref{eq:christ-34} are satisfied, and the signs of three nonzero stiffness coefficients composing the triplet $\big[ c_{1 5}, \, c_{3 5}, \, c_{4 6} \big]$ coincide, 
\begin{subnumcases}{\label{eq:christ-44}}
\big[ c_{1 5}, ~ c_{3 5}, ~ c_{4 6} \big] > 0 \, \text{\, or} & \\
\big[ c_{1 5}, ~ c_{3 5}, ~ c_{4 6} \big] < 0 \, . & 
\end{subnumcases}
Then equation~\ref{eq:christ-33}a, now in the form
\begin{equation} \label{eq:christ-45}
c_{1 5} \, p_1^2 + c_{4 6} \, p_2^2 \, + c_{3 5} \, p_3^2 = 0 \, ,
\end{equation}
has only trivial real-valued solution 
\begin{equation} \label{eq:christ-45.1}
\bm{p} \equiv [\, p_1, ~ p_2, ~ p_3] = \bm{0} \, ,
\end{equation}
making system~\ref{eq:christ-33} incompatible. 

As a result, our polarization direction $\bm{U} = [0, \, 0, \, 1]$ is unattainable to any plane wave regardless of its wavefront normal $\bm{n}$; and the field of polarization vectors $\bm{U}$ develops a \emph{hole} at the vertical. The size of this hole, outlining a solid angle of prohibited polarization directions, is \emph{finite} rather than infinitesimal because a small  finite perturbation of order $\varepsilon \ll 1$ of the components of vector $\bm{U} = [0, \, 0, \, 1]$ replaces equation~\ref{eq:christ-45} by
\begin{equation} \label{eq:christ-46}
c_{1 5} \, p_1^2 + c_{4 6} \, p_2^2 \, + c_{3 5} \, p_3^2 = \mathcal{O}(\varepsilon) \, 
\end{equation}
and solution~\ref{eq:christ-45.1} by a real-valued slowness vector that has the length 
\begin{equation} \label{eq:christ-47}
|\bm{p}| \sim \mathcal{O}(\sqrt{\varepsilon}) \, ,
\end{equation}
provided that the sign of the right side of equation~\ref{eq:christ-46} coincides with that of the stiffnesses in the triplet $\big[ c_{1 5}, \, c_{3 5}, \, c_{4 6} \big]$; otherwise, a solution becomes complex-valued. Clearly, the length of vector $\bm{p}$ given by relationship~\ref{eq:christ-47} is too small to ensure the compatibility of equations~\ref{eq:christ-33}.

\begin{figure}[t]
	\centering
	\includegraphics[width=0.45\textwidth]{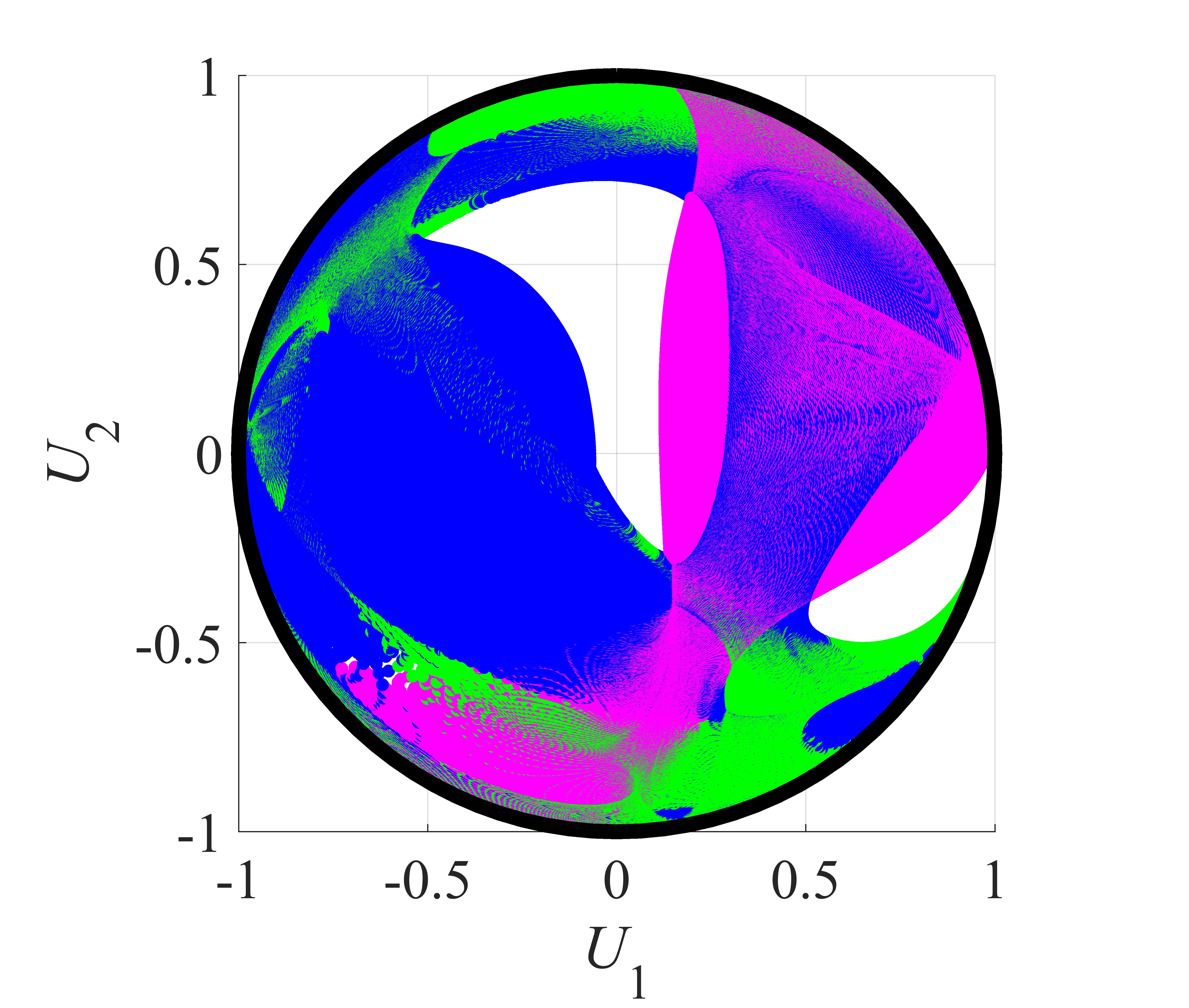}
	\caption{Projections of polarization vectors $\bm{U}^\mathrm{P}$ (magenta), $\bm{U}^\mathrm{S1}$ (blue), and $\bm{U}^\mathrm{S2}$ (green) of the P-, fast S$_1$-, and slow S$_2$-waves, respectively, onto the horizontal plane. The computations are carried out in model~\ref{eq:christ-49} for the wavefront normal directions $\bm{n}$, covering the entire unit sphere. 
	}
	\label{fig:U2p-07}
\end{figure}

To illustrate possible sizes of solid angles of prohibited polarization directions, we construct a triclinic model 
\begin{equation} \label{eq:christ-49}
\bm{c} = \left( \begin{array}{c c r r r r}
50 & 10 & \textcolor{blue}{\bm{-20}} & \bm{-5} & \textcolor{green}{\bm{1}} & 10 \\
~ &  50 &  20 &  3 & 6 & 10 \\
~ &  ~  & 50  &  0 & \textcolor{green}{\bm{4}} & \textcolor{red}{\bm{2}} \\
~ & \mathrm{SYM} & ~ & 22 & \textcolor{red}{\bm{-2}} & \textcolor{green}{\bm{5}} \\
~ &  ~ &  ~ &           ~ & \textcolor{blue}{\bm{20}} & \bm{5} \\
~ &  ~ &  ~ &           ~ & ~ & 10 \\
\end{array} \right) \! , 
\end{equation}
in which the boldface stiffness coefficients obeying conditions~\ref{eq:christ-34}a, \ref{eq:christ-34}b, \ref{eq:christ-34}c, and \ref{eq:christ-44}a are typeset in black, blue, red, and green, respectively. Next, we solve the Christoffel equation
\begin{equation} \label{eq:christ-02}
\bm{\Gamma}(\bm{n}) \bm{\cdot U} = V^2 \, \bm{U} \, ,
\end{equation}
analogous to equation~\ref{eq:christ-03}, in model~\ref{eq:christ-49} for a set of wavefront normals $\bm{n}$ spanning the entire unit sphere. The obtained polarization vectors $\bm{U}(\bm{n})$, displayed in Figure~\ref{fig:U2p-07}, exhibit two finite-size solid angles of prohibited polarization directions, appearing as the white areas.

\section{Discussion and conclusions}

The presented analysis allows us to list all the possibilities of describing plane-wave propagation in homogeneous anisotropic media in terms of a single input quantity~| the unit vector $\bm{n}$, $\bm{r}$, or $\bm{U}$.  

\begin{description}  \vspace{-2mm}
	\item[1. Input wavefront normal $\bm{n}$.] Solving the Christoffel equation~\ref{eq:christ-02} always yields three phase velocities $V_\mathrm{P}(\bm{n})$, $V_\mathrm{S1}(\bm{n})$, and $V_\mathrm{S2}(\bm{n})$ of the P-, S$_1$-, and S$_2$-waves, equal to square roots of the eigenvalues of Christoffel tensor $\bm{\Gamma}(\bm{n})$. 
	\begin{itemize}
		\item If $V_\mathrm{P}(\bm{n}) \neq V_\mathrm{S1}(\bm{n}) \neq V_\mathrm{S2}(\bm{n})$, the wavefront normal direction $\bm{n}$ is non-singular, and equation~\ref{eq:christ-02} results in three distinct eigenvectors~| the unit polarization vectors $\bm{U}^\mathrm{P}(\bm{n})$, $\bm{U}^\mathrm{S1}(\bm{n})$, and $\bm{U}^\mathrm{S2}(\bm{n})$. Then equation~\ref{eq:ani101-02} defines three group-velocity vectors $\bm{g}^\mathrm{P}(\bm{n})$, $\bm{g}^\mathrm{S1}(\bm{n})$, and $\bm{g}^\mathrm{S2}(\bm{n})$ and three unit ray-direction vectors $\bm{r}^\mathrm{P}(\bm{n})$, $\bm{r}^\mathrm{S1}(\bm{n})$, and $\bm{r}^\mathrm{S2}(\bm{n})$,  the triples of both vectors $\bm{g}(\bm{n})$ and $\bm{r}(\bm{n})$ uniquely determined.
		
		\item Alternatively, if at least two  phase velocities $V_\mathrm{P}(\bm{n})$, $V_\mathrm{S1}(\bm{n})$, and $V_\mathrm{S2}(\bm{n})$ coincide, the wavefront normal direction $\bm{n}$ becomes singular, $\bm{n} = \bm{n}^s$, 
		generally resulting in an infinite number of polarization vectors $\bm{U}$ of body waves corresponding to the singularity. Only at kiss singularities \cite[e.g.,][]{CrampinYedlin1981} the uniqueness of vectors $\bm{g}(\bm{n}^s)$ and $\bm{r}(\bm{n}^s)$ is maintained despite the nonuniqueness of $\bm{U}(\bm{n}^s)$. All other singularities
		produce an infinite number of vectors $\bm{g}(\bm{n}^s)$ and $\bm{r}(\bm{n}^s)$.
	\end{itemize} \vspace{-2mm}
	
	\item[2. Input ray direction $\bm{r}$.] Equations~22 in Grechka \cite{Grechka2017R2G}  or equations~2.C.13 and~2.C.14 in Grechka and Heigl\cite{GrechkaHeigl2017} comprise an algebraic system of degree~43 that has an odd number of non-centrally symmetric real-valued solutions $\big\{\bm{n}(\bm{r}), \, \bm{U}(\bm{r}) \big\}$, ranging from 3 to 19 and representing the uniquely determinable parameters of plane waves propagating along a given ray direction~$\bm{r}$. Although the system degenerates at kiss singularities because of the nonuniqueness of $\bm{U}(\bm{r})$, it yields a unique wavefront normal $\bm{n}(\bm{r}) = \bm{r}$ there.
	
	\item[3. Input polarization vector $\bm{U}$.] Equations~\ref{eq:christ-03} solved for the slowness vector $\bm{p}(\bm{U})$ can have from 0 to 4 or infinite number of real-valued solutions and the same number of solutions for the group-velocity vector $\bm{g}(\bm{p}, \, \bm{U})$.
\end{description} \vspace{-2mm}

Clearly, different inputs~| $\bm{n}$, $\bm{r}$, or $\bm{U}$~| lead to very different descriptions of wave propagation in homogeneous anisotropic media. The knowledge of either the wavefront normal $\bm{n}$ or the ray direction $\bm{r}$ guarantees the presence of at least three plane body-wave solutions, whereas the specification of polarization vector $\bm{U}$ can result in no solutions.




\vspace{-3mm}
\bibliography{refs-vg-arXiv}

\end{document}